\documentstyle[12pt]{article}

\begin{document}

\begin{flushright} 
Utrecht THU-96/18\\ 
gr-qc/9604003\\ 
\end{flushright} 
\vskip 1cm 
\begin{center}
{\LARGE\bf 1+1 Sector of 3+1 Gravity}\\ 
\vskip 1cm 
{\large Ted Jacobson} 
\vskip .5cm 
{\it Institute for Theoretical Physics, University of Utrecht\\
P.O. Box 80.006, 3508 TA Utrecht, The Netherlands}\\ 
and\\ {\it Department of Physics, University of Maryland\\
College Park, MD 20742-4111, USA}\\
	     {\tt jacobson@umdhep.umd.edu}

\end{center} \vskip 1cm

\begin{abstract}
The rank--1 sector of classical Ashtekar gravity is considered,
motivated by the degeneracy of the metric along the Wilson lines in
quantum loop states.  It is found that the lines behave like 1+1
dimensional spacetimes with a pair of massless complex fields 
propagating along them. The inclusion of matter and extension to 
supergravity are also considered. 
\end{abstract}

\def\E{E}
\def\beq{\begin{equation}}
\def\eeq{\end{equation}}
\def\D{{\cal D}}
\def\t{{\tau}}

In Ashtekar's Hamiltonian formulation of general relativity\cite{Asht}
the phase space variables $(\E^{ai}, A_{ai})$ are the same as in a
complexified $SO(3)$ Yang-Mills theory. $A_{ai}$ is a complex $SO(3)$
gauge field, and $\E^{ai}$ is a real triad of vector densities.  The
usual metric interpretation of general relativity requires the
``electric fields" $\E^{ai}$ to form a non-degenerate $3\times3$ matrix
so that the spatial metric defined by by $qq^{ab}=\E^{ai}\E^{bi}$ will
be non-degenerate.  (Here $q$ is the determinant of the metric, and the
$SO(3)$ indices are raised and lowered with the Kronecker delta.)
Nevertheless, Ashtekar's formalism remains well defined when $\E^{ai}$
is degenerate, so in fact it defines a particular degenerate extension
of general relativity.  This letter is concerned with that extension.

Much work on quantum general relativity has made use of this degenerate
sector of Ashtekar gravity\cite{JacoSmol,Abook}. In quantum ``loop"
states based on Wilson lines, the ``electric field" vanishes off the
lines and has rank 1 on the lines, where it takes the form
\beq
\E^{ai}=V^a\tau^i
\label{rank1}
\eeq
where $V^a$ is a vector tangent to the line and $\t^i$ is a vector in
the Lie algebra of $SO(3)$ that is a spatial density of unit weight.  
(At vertices where such lines meet $\E^{ai}$ may be non-degenerate.)

Until recently no spacetime interpretation for these degenerate
geometries was evident, but thanks to the recent work of
Matschull\cite{Mats} we now know that degenerate $\E^{ai}$'s yield a
definite causal structure for spacetime. For example, when $\E^{ai}=0$
the causal cone collapses to a line, while for the rank--1 form
(\ref{rank1}) it is a two dimensional wedge, as in 1+1 dimensional
Minkowski spacetime.\footnote{Bengtsson\cite{Beng1} showed that the
four dimensionally covariant theory defined by the action
$S[\theta,\omega]=\int d^4x\, \tilde{\theta}^{\mu A}\tilde{\theta}^{\mu
A} {}^{(+)}\Omega_{\mu\nu AB}$ is equivalent to the degenerate
extension of Ashtekar gravity, where ${}^{(+)}\Omega_{\mu\nu AB}$ is
the curvature of a self-dual $SO(3,1)$ connection and
$\tilde{\theta}^{\mu A}$ is a tetrad density of weight 1/2.
Matschull's work showed that the nonvanishing lapse condition in the
canonical theory is equivalent to the requirement that
$\tilde{\theta}^{\mu A}$ defines a (possibly degenerate) causal
structure.  This formulation will not be used here.} This suggests that
the loop states can be given an interpretation in which the lines
define 1+1 dimensional spacetimes that communicate at the vertices,
much as string worldsheets do when they split and join.

As a way of exploring this idea, the rank--1 sector of {\it classical}
Ashtekar gravity will be considered here.\footnote{Degenerate 
classical Ashtekar gravity has been explored in the 
past\cite{Beng1,Beng2,Beng3,JacoRoma,Roma,BMS,Reis}, 
although the full dynamics of the rank--1 sector seems not to have been previously determined.} The field will be taken here
to be smooth functions rather than distributions as in the loop
states. What we find
is that the 1+1 dimensional world is populated with a pair of
propagating massless complex fields, which are in fact the transverse
components of the gauge field. The field equation takes the form of a
pair of Dirac equations (although since the gauge field is bosonic the
significance of this is questionable) in a constant electrostatic
potential due to the presence of holonomy on closed loops. 
A disturbing feature is that the ``electrostatic potential"  
is in general complex, leading to solutions that grow exponentially 
in time. After treating the case of
classical general relativity in vacuum, the addition of matter,
supergravity, and quantization will be discussed briefly.

General relativity in the Ashtekar formulation is a constrained
Hamiltonian system. The phase space variables have Poisson brackets
given by $\{E^{ai}(x), A_{bj}(y)\}=i\delta^i_j\delta^b_a\delta^3(x,y)$.
Note that the electric field---the momentum conjugate to the gauge field
or connection---is a vector density of unit weight. Associated with the
gauge field $A_{ai}$ we have a covariant derivative $\D_a$ and curvature
$F_{ab}^i=\partial_a A_b^i-\partial_b A_a^i +\epsilon^{ijk}
A_{aj}A_{bk}$. The constraints are given by
\begin{eqnarray}
{\cal G}^i&:=&\D_a\E^{ai}=0\label{G}\\
{\cal V}_a&:=&\E^b_i F_{ab}=0\label{V}\\
{\cal S}&:=&\epsilon_{ijk}\E^{ai}\E^{bj}F_{ab}^k=0.\label{S}
\end{eqnarray}
These are called the Gauss, vector, and scalar constraints.  Initial
data is required to satisfy these constraints, and they are preserved
by evolution generated by the Hamiltonian which is an arbitrary linear
combination of the constraints (with nonvanishing coefficient
of ${\cal S}$).

The gauge field is complex, in fact it is the spatial restriction of
the anti-self-dual part of the spacetime spin connection in
non-degenerate solutions. The reality condition which relates the real
part of $A_{ai}$ to the spin connection determined by $\E^{ai}$ does
not make sense for degenerate triads.  However this is equivalent in
the nondegenerate case to
 the conditions that the inverse density weighted metric
$\E^{ai}\E^{bi}$ and its time derivative be
real\cite{ART}. These conditions continue to make sense in the
degenerate case and we shall take them as the reality conditions for
the degenerate extension. We also assume that $\E^{ai}\E^{bi}$ is
positive semi-definite.  For $\E^{ai}$ of the form
(\ref{rank1}) this implies that $V^a$  can be chosen
to be real, with $\t^i\t^i$ real and positive. 
Since it turns out that $\E^{ai}$ of this form is constant
in time, all these reality conditions will therefore be satisfied.

Now let us ask if there is an autonomous sector of this theory in which
$\E^{ai}$ takes the rank--1 form (\ref{rank1}) for some vector field
$V^a$ and $SO(3)$ vector density $\t^i$. First note that the scalar
constraint ${\cal S}$ (\ref{S}) vanishes identically with this ansatz.
The remainder of the analysis is made more transparent if at this stage
some of the gauge freedom is fixed.  To begin with the gauge symmetry
consists of local $SO(3)$ rotations, spatial diffeomorphisms, and local
time reparametrizations. Let us partly fix the spatial diffeomorphisms
by choosing coordinates $(x^\alpha,z)$, $\alpha=1,2$, so that the
integral curves of $V^a$ are the lines of constant $x^\alpha$, and the
components of $V^a$ are just $(0,0,1)$. That is, 
$V^a=(\partial/\partial z)^a$.
In making this choice we are
either restricting to a neighborhood (which could be the whole space)
in which the curves form a nice congruence that does not wrap back
densely on itself. With these coordinates, the Gauss and vector
constraints become
\begin{eqnarray}
\partial_z\t^i+\epsilon^{ijk}A_{zj}\t_k&=&0\\
\t_i F_{z\alpha}^i&=&0
\end{eqnarray}
The Gauss constraint thus states that the $SO(3)$ vector $\t^i$ is parallel
transported along the curves. If a curve forms a closed loop then
$\t^i$ must therefore be parallel to the logarithm of the
holonomy element for that
loop. We can thus adopt a gauge with $A_{zi}$ constant on each line,
\beq
A_z^i=H(x^\alpha)\t^i
\label{gauge}
\eeq 
(assuming, as we shall do
here, that the $SO(3)$ bundle admits a global product structure). 

For the moment let us for simplicity assume there are no closed orbits of
$V^a$. In this case, the $SO(3)$ gauge can always be chosen so that
$A_{zi}=0$, and the constraints further simplify to 
\begin{eqnarray}
\partial_z\t^i&=&0\\
\t^i \partial_zA_{\alpha i}&=&0\label{8}
\end{eqnarray}
The Gauss law thus implies that $\t^i$ is independent of $z$, and the
vector constraint implies that the $\t^i$-component of $A_\alpha^i$ is
independent of $z$.

We have partly fixed the gauge freedom and solved the constraints.  The
next task is to examine the equations of motion---Hamilton's
equations.  The general Hamiltonian is a linear combination of the
constraints, 
$H=\int d^3x\, (-{1\over2}N{\cal S}+iN^a{\cal V}_a+i\Lambda^i{\cal G}_i)$.
However, not all choices are consistent with the gauge choices already
made. A consistent choice is $N=1$, $N^a=0$, $\Lambda^i=0$. For this
choice, the equations of motion imply $\partial_t\E^{ai}=0$ and
$\partial_t A_{zi}=0$, so the evolution is consistent with the rank--1
ansatz and the gauge choice. The only nontrivial dynamics occurs for
the transverse components $A_\alpha^i$ of the gauge field, which
satisfy
\begin{eqnarray}
\partial_t A_\alpha^i&=&i\epsilon^{ijk}\t^jF_{\alpha z}{}^k\label{9}\\
&=&-i\epsilon^{ijk}\t^j\partial_z A_\alpha{}^k.\label{eom}
\end{eqnarray}
The $\t_i$-component of $A_\alpha^i$ is thus time independent, and the
only propagating variables are the transverse-transverse components.

To best reveal the structure of this equation 
we choose an orthonormal basis for the $SO(3)$ vectors such that the 
components of $\t_i$ are $(0,0,\t)$,
where $\t:=(\t^i\t^i)^{1/2}$, and denote the transverse-transverse components  
$A_\alpha^I$, $I=1,2$.
The field equation (\ref{eom}) then takes the form 
\beq
\partial_t A_\alpha^I=i\t\epsilon^{IJ}\partial_z A_\alpha{}^J.
\label{Dirac}
\eeq
 This can be recognized as a pair of 1+1 dimensional massless Dirac
equations for Dirac ``spinors" $A_1^I$, $A_2^I$. The Dirac equation
reads $i\partial_t\psi=\alpha^zp_z\psi$, so we read off that
$\alpha^z=\t\sigma_2$, and the 1+1 spacetime metric is
$ds^2=dt^2-\t^{-2} dz^2$. Since $\t$ is constant, this metric is flat.
The equation of motion can be diagonalized to yield $\partial_t(A^1\pm
iA^2)=\pm\partial_z(A^1\pm iA^2)$. The complex linear combinations
$(A^1\pm iA^2)$ propagate in opposite directions along the curves at
the speed of light.

To complete the analysis of the classical theory the remaining gauge
freedom should be fixed to identify the physical degrees of freedom.
The remaining freedom consists of coordinate transformations in the
transverse ($\{x^\alpha\}$) subspace, and  $SO(3)$ rotations about the
$\t^i$-axis. The $\t^i$-component of the connection is a gauge field
for this $U(1)$ subgroup, and is required by the constraints and
equations of motion to be independent of $z$ and $t$. Finally, 
the transverse coordinates $x^\alpha$ can be chosen so that the scalar density $\t=1$. This leaves
unfixed the ``area" preserving diffeomorphisms in the transverse space
(thinking of $\t$ as an area element). Gauge fields
related by these transformations should presumably be identified.

In the remainder of this letter four topics will be discussed briefly:
allowing for holonomy on closed loops, addition of matter, extension to
supergravity, and quantization.

Let us now go back and allow for the possibility that there is
holonomy on closed loops. Then we cannot set $A_{zi}=0$,
but we can set $A_z^i=H(x^\alpha)\t^i$ for some function 
$H$ of the transverse coordinates. The only change this makes in the 
constraints is that the vector constraint no longer takes the form 
(\ref{8}), but rather  
\beq
\t^i \partial_zA_\alpha^i= \t^i \partial_\alpha H \t^i.
\label{holon}
\eeq
It would be rather strange if any real dependence on a transverse
derivative remained, since points on different curves are effectively
at infinite distance from each other in the degenerate metric.  In fact
the component $A_z^i$ can be used as a gauge parameter for a gauge
transformation which absorbs the new term on the right hand side of
(\ref{holon}) into the left hand side. In so doing, the transverse
$SO(3)$ gauge freedom is thus largely fixed. Turning to the equations
of motion, $\E^{ai}$ is still constant, while from (\ref{9}) we see
that $A_{zi}$ and $\t^iA_{\alpha i}$ are also constant. The only change
arises from the $A_zA_\alpha$--term in $F_{z\alpha}$, which modifies
the equation of motion (\ref{Dirac}). The new equation is
\beq
\partial_t A_\alpha^I=i\t\epsilon^{IJ}\partial_z A_\alpha{}^J
-i\t^2 H\, A_\alpha^I.
\label{heom}
\eeq
If $H$ is real 
the new holonomy term behaves like a constant electrostatic potential 
(which depends on $(x^\alpha)$ ) in the ``Dirac" equation and can 
be absorbed by a time-dependent phase transformation of $A_\alpha^I$. 
If $H$ is complex then the solutions to (\ref{heom}) grow exponentially
in one time direction, so there appears to be an instability. Our reality
conditions do not rule out this peculiar behavior, so perhaps there is 
something really sick about our degenerate theory.

Turning now to the addition of matter, the first consideration
is the constraints. Since the rank--1 ansatz for $\E^{ai}$ 
annihilates the gravitational scalar constraint  (\ref{S}) 
by itself, the matter contribution
to the scalar constraint must also vanish. For a scalar 
field $\varphi$ this takes the form\cite{ART} 
$\pi^2 + (\E^{ai}\partial_a \varphi)^2$,
so the only possibility is that $\pi$ vanishes and $\varphi$ is
constant along the lines. For a spinor field $\psi^A$ 
the contribution to the 
scalar constraint takes the form\cite{ART,Jaco-spin} 
$\pi^A\E^a{}_{AB}\D_a\psi^A$,
where we now employ a pair of symmetric 
spinor indices for the triad rather than a single $SO(3)$ index.
This contribution vanishes if $\pi^A=0$ (or, essentially
equivalently, if $\partial_z\psi^A=0$). This ansatz also kills the 
contributions to the vector and Gauss law constraints and is 
preserved by the time evolution. The time evolution of $\psi^A$
is then just the Weyl equation in the 1+1 dimensional worlds.
An important comment is that the non-polynomial 
reality condition relating
$\pi^A$ to the conjugate of $\psi^A$ is abandoned in this ansatz. 
However there is 
a polynomial form\cite{ART} of the reality conditions for a spinor field:
$(\E^a{}_{MN}\psi^M\pi^N)^*=(\E^a{}_{MN}\psi^M\pi^N)$ and
$(\E^a{}_{MN}\pi^M\pi^N)=({\rm det}\E)^2(\E^a{}_{MN}\psi^M\psi^N)$.
For our ansatz these conditions are satisfied.
Addition of mass terms or a cosmological constant term to the
constraints adds absolutely nothing because these and their
contributions to the equations of motion vanish identically
for the rank--1 ansatz. Also Yang-Mills fields cannot be sensibly
added, since to achieve polynomiality the scalar constraint must
be taken of density weight 2\cite{ART}. 
The resulting density contributes
nothing to the constraints or equations of motion with the rank--1
ansatz, and the gravitational terms are all killed as a result
of the multiplication by $({\rm det}\E)^2$.

The extension of the rank--1 ansatz to N=1 supergravity 
in the Ashtekar form\cite{Jaco-sg} will now be considered. 
The additional canonical variables are 
the gravitino field $\psi_{aM}$ and its conjugate momentum 
$\pi^{aM}$. Clearly the extension of the
rank--1 ansatz is 
\beq
\E^a{}_{MN}=V^a\t_{MN},\qquad \pi^{aM}=V^a \lambda^M.
\label{srank1}
\eeq
The right-handed supersymmetry constraint, 
${\cal S}^{\dagger A}=\E^{aAM}\E^{b}{}_{MN}\D_{[a}\psi_{b]}{}^N=0$,
is then identically satisfied, while the left handed 
supersymmetry constraint ${\cal S}^A=\D_a\pi^{aA}=0$ implies
that the spinor $\lambda^A$ is covariantly constant along the
lines. This puts an immediate restriction on the holonomy: 
{\it there can be none}, since the only $SL(2,C)$ element with unit
eigenvalues is the identity. From now on we therefore assume
the holonomy vanishes.\footnote{One would think that there should be no
problem supersymmetrizing the holonomy as well, so I suspect that there
is an improved ansatz that would allow for holonomy.}
 The left handed supersymmetry gauge can
then be chosen so that $V^a\psi_a^M=0$\footnote{Under this
symmetry generated by $\epsilon^M$ we have 
$\delta\psi_a{}^M\sim\D_a\epsilon^M$ and 
$\delta \E^{aMN}\sim \pi^{a(M}\epsilon^{N)}=V^a\lambda^{(M}
\epsilon^{N)}$.
Thus in achieving $V^a\psi_a{}^M=0$ we might be forced to 
include a Grassmanian part into $\t^{MN}$.}, and the $SL(2,C)$ gauge
can be chosen so that $V^aA_a^{MN}=0$. With these gauge choices
the left handed supersymmetry constraint implies that $\lambda^M$
is constant on the curves and
the Gauss law 
${\cal G}_{AB}=\D_a\E^a{}_{MN}+\pi^a{}_{(M}\psi_{M)a}=0$
reduces to the pure gravity case, which states that 
$\t_{MN}$ is constant along the curves. 

To complete the story the scalar and vector constraints and equations
of motion must be considered.  Here we face the question of {\it which}
degenerate extension of supergravity we wish to consider. It was found
in \cite{Jaco-sg} that the scalar and vector constraints do not take a
polynomial form when derived directly from the chiral action in
distinction with the pure gravity case. This form of the theory is
therefore not suitable for a degenerate extension. However, these
constraints do take a polynomial form when defined via the Poisson
bracket $\{{\cal S}^A,{\cal S}^{\dagger B}\}$. It is this form of the
constraints that will therefore be taken to define the theory. It is
simple to see that this bracket vanishes identically for the
supersymmetrized rank--1 ansatz (\ref{srank1}), so these constraints
are satisfied.\footnote{In the pure gravity case, the vector constraint
(\ref{8}) which demands that the $\t^i$ component of the transverse
connection is independent of $z$ is therefore missed in this degenerate
extension. The difference between these two degenerate extensions of
gravity was studied in Refs. \cite{JacoRoma,Roma,Mats-sg}.} 
The equations of motion with a simple choice of the
Lagrange multipliers follow from Poisson bracket with the Hamiltonian
$H=i\int N\{{\cal S}^A,{\cal S}^{\dagger}{}_A\}=
i\int N tr\Bigl(\E^aE^b F_{ab} +
4\pi^a \E^b \D_{[a}\psi_{b]}\Bigr)$, 
written here in a notation with the spinor indices suppressed. Under
this evolution the quantities $\E^{aMN}$, $\pi^{aM}$, $A_{zMN}$, and
$\psi_{zM}$ are all constant, in agreement with our ansatz and gauge
conditions.  The remaining field equations determine the evolution of
the transverse components $\psi_\alpha$ and $A_\alpha$.  We find (up to
constant factors that have not been worked out)
$\partial_t\psi_\alpha\sim i\t\partial_z\psi_\alpha$, so each
transverse component of $\psi_\alpha$ satisfies the 1+1 dimensional
Dirac equation, while the equation previously found for $A_\alpha$ in
the pure gravity case picks up an additional term linear in
$\psi_\alpha$:
\beq
\partial_t A_\alpha{}^{MN}\sim 
i\Bigl(\t^{P(M}\partial_zA_\alpha^{N)}{}_P
+\lambda^{(M}\partial_z\psi_\alpha{}^{N)}\Bigr).
\eeq
Another difference with the pure gravity case is that the 
$\t$--component of $A_\alpha$ now seems to have some nontrivial
dynamics, coming from the second term above.

Finally, let us just raise some questions about quantization.  Our
investigation was motivated by the existence of line-like excitations
in the quantum theory, however the propagating fields we have found on
the lines are not evident in these quantum loop states.\footnote{
Although our fields are smooth, whereas in the loop states the triad is
distributional, this smoothness does not seem critical in establishing
the existence of these propagating modes.} It seems these modes are in some
sense excitations around the Wilson lines. They involve the {\it
transverse} connection components, so perhaps they should be thought of
as pertaining to correlations between the different Wilson lines. 
How should these appear in the connection or loop representations?
Perhaps they correspond to the presence of intersections (vertices)
in network states. Classically,
these excitations behave as free field theories in 1+1
dimensions. From all the work on conformal field theories we expect
some nontrivial consistency conditions to emerge when these fields are
quantized.  Are such consistency conditions simply absent in the
nonperturbative quantization schemes that have been explored, or are
they lurking around some corner waiting to be discovered?
Last, but not least, what is the quantum incarnation of the instability
found in the classical theory in the presence of imaginary holonomy?

\section*{Acknowledgments}
I am grateful to  H.J. Matschull for discussions and to J. Lewandowski
for encouragement and critical comments that saved me from some errors. 
This work was supported in part by NSF Grant
PHY94-13253, the University of Utrecht, and a University of Maryland GRB
award.

\end{document}